\documentclass[10pt,journal]{IEEEtran}
\usepackage[font=small,labelsep=space]{caption}
\usepackage{verbatim}
\usepackage{graphicx}
\usepackage{subfig}
\usepackage{multirow}
\usepackage{amssymb}
\usepackage{amsmath}
\usepackage{amsthm}
\usepackage{amsfonts}
\usepackage{paralist}
\usepackage{url}
\usepackage{epstopdf}
\usepackage{float}
\usepackage[justification=centering]{caption}
\usepackage{color}
\usepackage{paralist}
\usepackage[noend]{algpseudocode}
\usepackage{algorithmicx}
\usepackage{algorithm}
\let\itemize\compactitem
\let\enditemize\endcompactitem
\let\enumerate\compactenum
\let\endenumerate\endcompactenum

\begin{document}

\title{A Design of SDR-based Pseudo-Analog Wireless Video Transmission System}

\author{Xiao-Wei~Tang,~\IEEEmembership{Student Member,~IEEE}, Xin-Lin~Huang*,~\IEEEmembership{Senior Member,~IEEE}, \\

\thanks{Xiao-Wei Tang (email: {\tt xwtang@tongji.edu.cn}) is with the Department of Control Science and Engineering, Tongji University, Shanghai 201804, China.}
\thanks{Xin-Lin Huang (email: {\tt xlhitcrc@163.com}) is with the Department of Information and Communication Engineering, Tongji University, Shanghai 201804, China (corresponding author).}
}

\markboth{}{Tang \MakeLowercase{\textit{et al.}}: A Design of SDR-based Pseudo-Analog Wireless Video Transmission System \ldots}
\maketitle

\begin{abstract}
The pseudo-analog wireless video transmission technology can improve the effectiveness, reliability, and robustness of the conventional digital system in video broadcast scenarios. Although some prototypes of IEEE 802.11 series have been developed for researchers to do simulations and experiments, they are usually expensive and provide very limited access to the physical layer. More importantly, these prototypes cannot be used to verify the correctness of the new proposed pseudo-analog wireless video transmission algorithms directly due to limited modulation modes they can support. In this paper, we present a novel design of software radio platform (SDR)-based pseudo-analog wireless video transceiver which is completely transparent and allows users to learn all the implementation details. Firstly, we prove that the analog method can also achieve the optimal performance as the digital method from the perspective of the rate-distortion theory. Then, we describe the two hardware implementation difficulties existed in the designing process including the data format modification and the non-linear distortion. Next, we introduce the implementation details of the designed transceiver. Finally, we analyze the performance of the designed transceiver. Specifically, the results show that the designed system can work effectively in both simulations and experiments.
\end{abstract}
\begin{IEEEkeywords}
SDR, Pseudo analog transmission, GNU Radio, USRP.
\end{IEEEkeywords}

\IEEEpeerreviewmaketitle

\section{Introduction}
\label{intro}
The software radio (SDR) platform is a programmable radio that allows users to access all data in the form of physical waveforms and to implement all signal processing steps on software [1]. This feature makes the SDR platform particularly suitable for developing early transceivers which can be used to verify the effectiveness of new proposed algorithms. More importantly, the open source SDR platform can switch between simulations and experiments seamlessly, which helps bridge the gap between theory and practice and reveal potential defects in the system design.

Although many companies, e.g., Cohda Wireless, NEC, and Denso, have developed prototypes of IEEE 802.11 series [2], these prototypes are usually expensive and provide limited access to the physical layer. Differently, the completely transparent SDR-based transceiver can allow users to access all the implementation details and even to modify the details when necessary, whose general structure is presented in Fig. 1. There are mainly eight modules in the SDR-based transceiver [3] including: 1) the communication module (e.g., SDR platform), 2) the driver module (e.g., universal software radio peripheral (USRP) hardware driver (UHD)), 3) the operating system module (e.g., system library, system interface, and kernel), 4) the interface module (e.g., universal serial bus (USB)/peripheral component interconnect express (PCIe)), 5) the control module (e.g., receive control), 6) the conversion module (e.g., digital up converter (DUC), digital down converter (DDC), digital-to-analog converter (DAC), and analog-to-digital converter (ADC)), 7) the filter module (e.g., filter), and 8) the power amplifier module (e.g., power amplifier).
\begin{figure}[htbp!]
\centering
\includegraphics[width=0.45\textwidth]{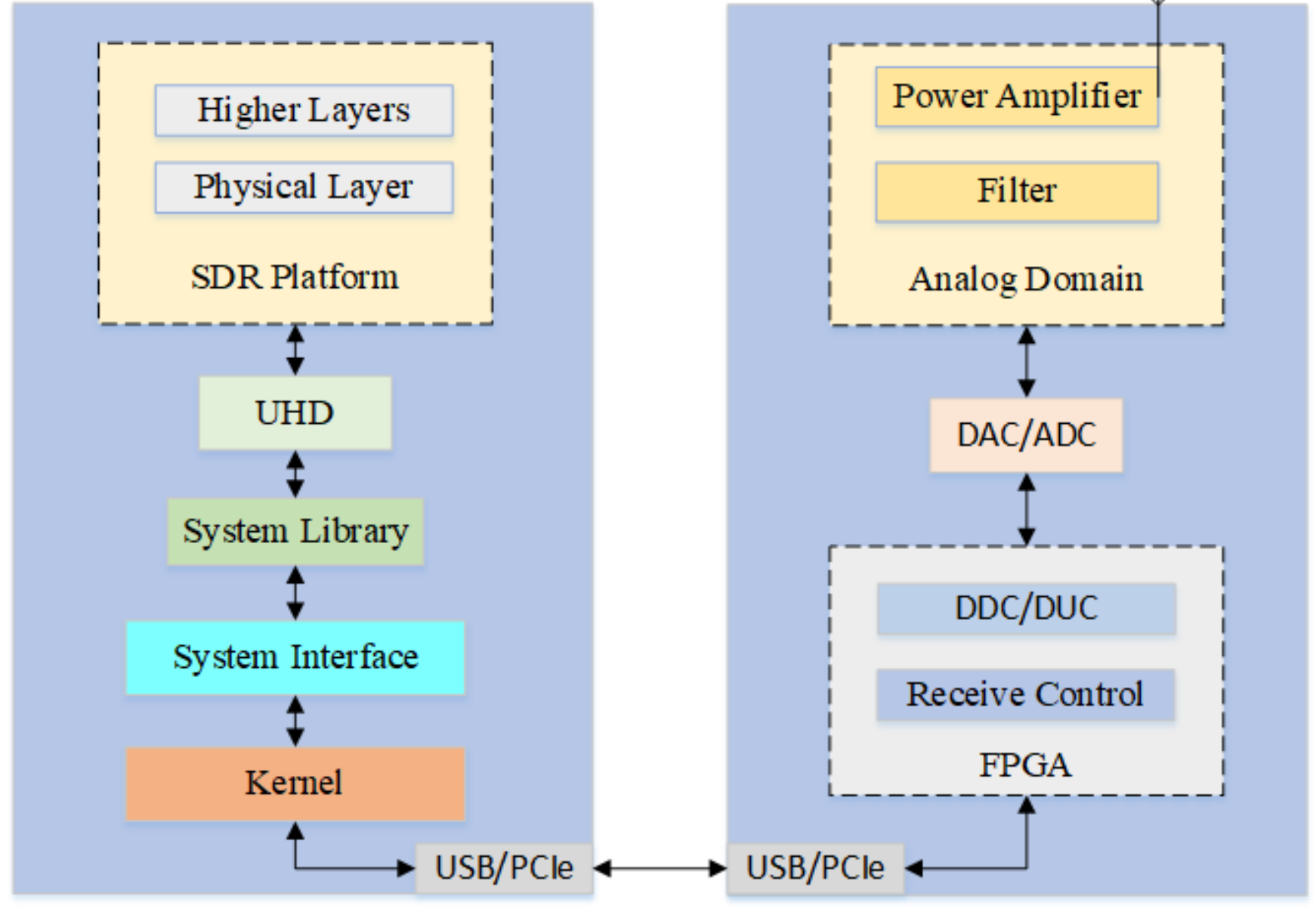}
\caption{The general structure of SDR-based wireless transceivers.}
\label{F1}
\end{figure}

Existing SDR platforms can be divided into two different categories according to the implementation type of physical layer including field programmable gate array (FPGA)-based SDR platforms and general purpose processor (GPP)-based SDR platforms [4]. The wireless open-access research platform (WARP) [5] and the LabVIEW-based USRP-RIO series [6] are two popular FPGA-based SDR platforms. FPGA-based SDR platforms can meet the strict delay constraints of current communication standards owing to their characteristics of deterministic timing and low latency. However, the ultra-high licensing fees and low flexibility of the physical layer prevent the FPGA-based SDR platforms being widely used by researchers. Instead, GPP-based SDR platforms are especially suitable for rapid prototype development and physical layer simulations [7]. Specifically, GPP-based SDR platforms are usually written in C$++$ or Python and can be implemented on personal computers (PCs), while the FPGA-based SDR platforms often use very high speed integrated circuit hardware description language (VHDL) or Verilog hardware description language which is hard to write, compile and test.

Sora [8] and GNU Radio [9] are two popular GPP-based SDR platforms, which can be connected to USRP devices provided by Ettus Research [10]. Sora is a fully programmable and high-performance SDR platform that can be used to implement most advanced wireless communication technologies [11]. GNU Radio provides a graphical interface to configure and debug transceivers, which can display physical waveforms of signals in real time. In addition, GNU Radio can be used on a variety of operating systems such as MacOS and Linux. Compared with Sora, GNU Radio's simple and practical graphical programming interface makes it a preferred choice for researchers.

In GNU Radio, all modules are designed based on the concept of unlimited data flow. There are mainly two data streaming mechanisms in GUN Radio including 1) message mechanism and 2) tag mechanism. In the message mechanism, protocol data units (PDUs) are transferred between modules in asynchronous mode. In addition, the message interfaces provide a convenient communication connection between external applications and GNU Radio modules. The main disadvantages of message mechanism are that it works asynchronously and the reliability of message transmission between modules can not be guaranteed. In tag mechanism, one tag stream is transmitted in parallel with the data stream. Unlike the data stream, the tag stream is used to transmit control information and data-related information. The tags can help modules identify the sampling points that need to be processed, thus enhancing the reliability of inter module transmission.

In this paper, we present a design of SDR-based pseudo-analog wireless video transmission system. The designed system can be used for both simulations and experiments, thus providing seamless switch between theory and practice. Specifically, we redefine the data format transmitted in the physical layer while retaining most of the original functions of the orthogonal frequency division multiplexing (OFDM) system (e.g., channel estimation and channel equalization). Compared with the conventional digital transmission framework (e.g., IEEE 802.11a), the designed system removes the quantization module and the channel coding module in the transmitter, and directly transmits complex signals formed by real pixel values.

\textbf{Contributions}:
\begin{enumerate}
\item A novel design of SDR-based pseudo-analog wireless video transmission system is proposed in which the graphical interface can greatly reduce the operation difficulty. It provides a convenient and effective experimental platform for the research of pseudo-analog video transmission technologies, providing a seamless switch between simulations and experiments.
\item From the perspective of Shannon theory, we reveal the reason of the inherent performance loss caused by the separate source and channel coding which is adopted in the conventional digital transmission system. In addition, we explain that the analog method can also obtain the same optimal performance as digital method from the perspective of rate-distortion theory.
\end{enumerate}

The rest of the paper is organized as follows. Section II introduces the related work including the rate-distortion theory and the pseudo-analog wireless transmission technology. In Section III, we analyze the difficulties of the design and implementation of the proposed system. Section IV gives the details of the designed system including the overview of the system and the detailed functions of each module in the order of transmitter, channel, and receiver. In Section V, we provide the implementation details and analyze the simulation and experiment results obtained by the designed system. Finally, we summarize this paper in Section VI.

\section{Related Work}
\label{sec:1}
In this section, we will introduce the rate-distortion theory and the existing pseudo-analog transmission systems. Specifically, we will firstly prove that the analog method can also achieve the optimal performance as digital method from the perspective of the rate-distortion theory. Then, we will introduce the difference between the pseudo-analog transmission system and the existing digital system, as well as the development of the pseudo-analog transmission systems.
\subsection{Rate-Distortion Theory}
\label{sec:2}
Shannon theory indicates that separate source and channel coding can achieve asymptotically optimal performance in the point-to-point communication [12]. Such conclusion promotes the design of layered communication system, where the physical layer is responsible for processing channel characteristic (e.g., path loss and fading) so that the application layer does not need to deal with bit errors. On the contrary, the application layer handles domain-specific issues (e.g., quantization and coding), making the physical layer invisible to the transmitted data. Although such a separate structure greatly reduces the complexity of the overall system design, it also destroys the essential relationships between the transmitted bitstream and the original pixels. Denote a memoryless Gaussian source as $X$ and its distribution as $X\!\!\sim\!{\cal N}\left( {0,\lambda } \right)$ where $\lambda$ represents the variance of $X$. Consequently, the corresponding rate-distortion function can be denoted as follows.
\begin{equation}
R(D_d) = \left\{ \begin{array}{l}
\frac{1}{2}\log \frac{\lambda }{D_d},\;\;0 \le D_d \le \lambda, \\
0,\;\;\;\;\;\;\;\;\;\;\;\;D_d > \;\lambda, \;\;\;\;
\end{array} \right.
\end{equation}
where $D$ denotes the distortion caused in the transmission process. $R(D_d)$ denotes the minimum source coding rate required to achieve the distortion $D_d$. According to the capacity of additive white Gaussian noise (AWGN) channel, i.e., $C = \log (1 + \gamma)/2$ and the rate-distortion formula given in Eqn.(1), the minimum distortion of digital communication can be denoted as follows.
\begin{equation}
{D_{d}} = \frac{\lambda }{{1 + \gamma}},
\end{equation}
where $\gamma$ represents the channel signal-to-noise ratio (SNR). Note that Shannon theory is actually idealized because it assumes that the channel condition and the source distribution remain unchanged, and the length of the source code is infinite. However, the physical layer can only process signals with limited length in practice. Therefore, in some specific communication scenarios (e.g., wireless video broadcast), such separate source and channel coding system may result in inherent performance loss. Fortunately, Goblick et al. have proved that linear analog coding can also achieve the optimal performance [13]. Consider the case of transmitting the source $X$ in AWGN channel using analog coding, that is, transmitting the source $X$ directly after power scaling. Assuming that the total transmission power is $P$ and the channel noise power is $\sigma _n^2$. In order to minimize the distortion, the transmitter assigns a scaling factor, i.e., $G = \sqrt {P/\lambda }$, for the source $X$. Therefore, the received signals can be represented as follows.
\begin{equation}
Y = GX + W,
\end{equation}
where $W$ denotes the Gaussian noise. According to the minimum mean squared error (MMSE) estimation criterion, the estimation of $X$ can be denoted as
\begin{equation}
{X_{M\!M\!S\!E}} = \frac{G\lambda}{{P + \sigma _n^2}}Y.
\end{equation}
Therefore, the distortion using the analog method can be calculated as follows
\begin{equation}
\begin{array}{l}
D_a = E{({X_{M\!M\!S\!E}} - X)^2}\\
\;\;\;\;\;\; = E{(\frac{G\lambda}{{P + \sigma _n^2}}Y - X)^2}\\
\;\;\;\;\;\; = \frac{\lambda}{1 + \gamma},
\end{array}
\end{equation}
where $E(\cdot)$ represents the expectation. Comparing Eqn. (5) with Eqn. (2), one can conclude that the analog method can also achieve the same optimal performance as the digital method.

\subsection{Pseudo-Analog Wireless Transmission Technology}
Conventional digital wireless video transmission systems often divide a video into several groups of pictures (GOPs) via standard encoder e.g., JPEG 2000 and H.264, and adopt predictive source coding [14-19]. Intra- and inter-frame correlations of the video enable high compression efficiency. The digital communication system can select a suitable channel modulation/coding scheme (MCS) to overcome the channel interference. However, the selected MCS may not guarantee a predetermined packet loss ratio (PLR) due to the fluctuations of the channel quality. The cliff effect will occur under deep channel fading. Especially in the broadcast scenario, the receivers have different channel qualities. Therefore, the transmitter should select a MCS according to the worst channel quality to guarantee the correct demodulation of all receivers. Consequently, the existing digital wireless video transmission system is ineffective in the broadcast scenario. The pseudo-analog transmission technology has been proposed to improve the effectiveness of the video broadcast [22-28]. In the pseudo-analog video broadcast system, the transmitter does not need to know the receivers' channel qualities because the receivers can demodulate the video adaptively according to their own channel qualities.

The first pseudo-analog video transmission system named SoftCast is proposed in [22]. At the transmitter, SoftCast first divides the original video into multiple GOPs. Next, SoftCast performs three dimensional discrete cosine transform (3D-DCT) for each GOP to remove the spatial and temporal redundancy. The transformed coefficients are divided into blocks with uniform size. The transmitter allocates different transmission power for each divided block according to its pixel variance. Then, Hadamard transform is performed for each block to reduce the peak-to-average power ratio (PAPR). Finally, the transmitter sends these transformed coefficients in high-density modulation mode. At each receiver, a series of operations are performed in order including the signal demodulation, inverse Hadamard transform, MMSE estimation, inverse 3D-DCT transform.

A large number of variants of SoftCast have also been proposed. In [23], a pseudo-analog transmission system called D-Cast is proposed where the correlations between video frames are fully utilized to enhance the video quality. Specifically, the received video frames are considered as the side information to assist with the reconstruction of the current frame. A data-assisted cloud radio access network for visual communications named DAC-RAN is proposed in [24]. DAC-RAN separates the control plane and the data plane in the conventional digital transmission infrastructure, and integrates a new data plane (that is specifically designed for video communications) into the virtual base station. The correlated information retrieved from video data is utilized as the prior knowledge in the video reconstruction. However, the quality of reconstructed video does not increase linearly with the increase of signal-to-noise ratio (SNR) due to mutual interference. Huang et al. propose a knowledge-enhanced wireless video transmission system named KMV-Cast [25-27] which could further eliminates the mutual interference. Liu et al. add the multiple-input multiple-output (MIMO) technique into SoftCast and point out that important coefficients should be transmitted on channels with large gain [28].

\section{Hardware Implementation Difficulties.}
In this section, we will describe two hardware implementation difficulties in the designed system including the data format modification and the non-linear distortion. Firstly, we will explain the difference between the conventional digital system and the designed system in terms of the coding and modulation scheme. Then, we will introduce the causes of non-linear distortion in the designed system and the countermeasures we take to reduce such non-linear distortion.

\subsection{Data Format Modification}
In the conventional digital system, the physical layer receives packets of bits from the application layer. These bits are first error protected by forward error correction (FEC) coding and interleaving, and then modulated into complex samples which are transmitted over the channel [29]. Fig. 2 shows two examples of the coding and modulation scheme in the conventional OFDM system and the designed OFDM system, respectively. In Fig. 2(a), the conventional OFDM system first adopts the FEC coding to increase the redundancy and then adopts the 16 quadrature amplitude modulation (16-QAM) to map each four data bits into a complex I/Q signal. Different from the conventional digital system, the physical layer of the designed system receives packets of real-value samples from the application layer. Specifically, these real-value samples are power-adjusted using the joint source-channel coding for error protection. The designed physical layer bypasses the FEC coding and interleaving modules, and directly maps each two real-value samples into a complex I/Q signal, as shown in Fig. 2(b).

In order to integrate such design into the standard communication system, we take advantage of the fact that the existing OFDM system separates the channel estimation and tracking from data transmission [30]. Therefore, we can change the way that the transmitted data is encoded and modulated without affecting the performance of OFDM. Specifically, the existing OFDM system divides spectrum into many independent subcarriers, some of which are used for channel tracking (the specific signals transmitted on these subcarriers are called pilots) while others are used for data transmission. Since the designed pseudo-analog video transmission system does not modify the pilots, the original OFDM functions including synchronization, carrier frequency offset estimation, channel estimation, and phase tracking will not be affected [31].
\begin{figure*}[htbp!]
\centering
\includegraphics[width=0.9\textwidth]{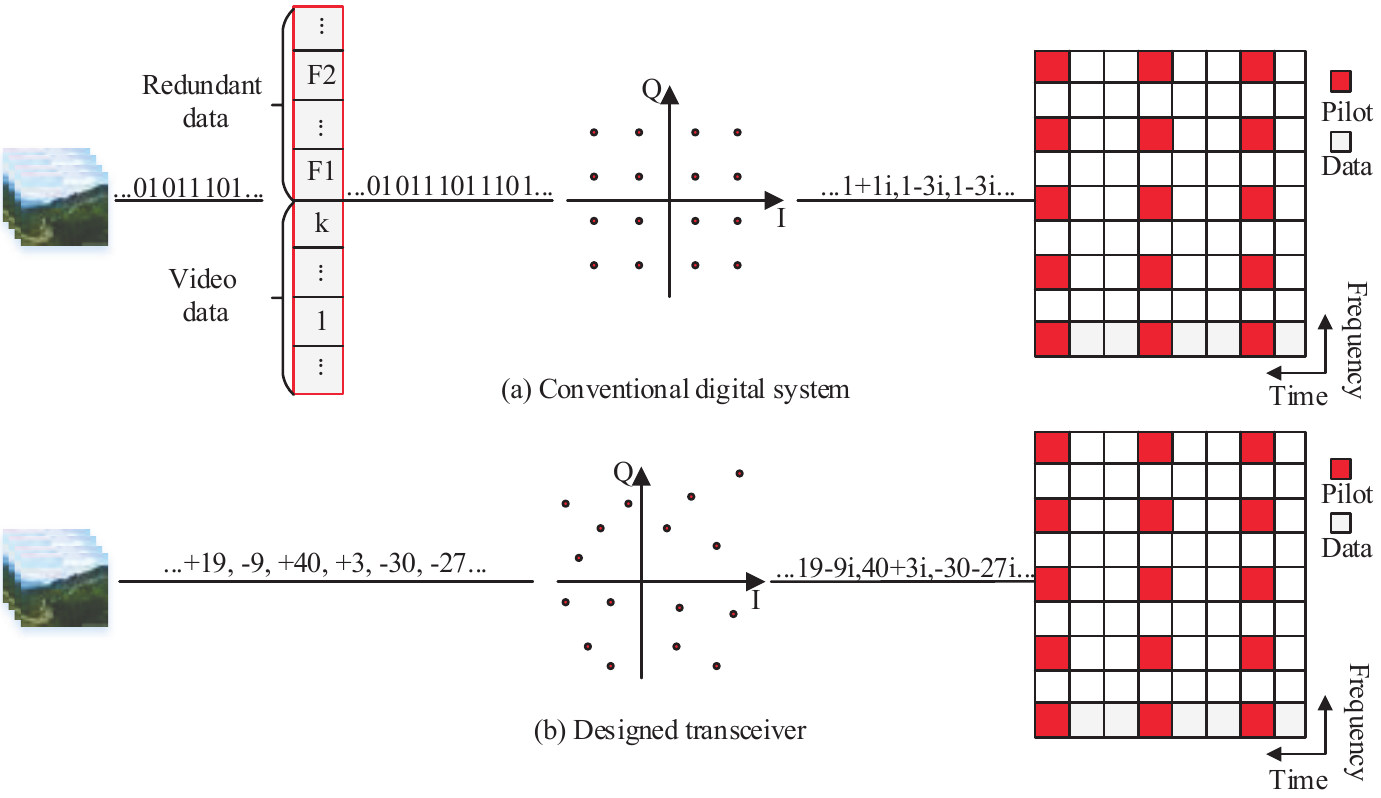}
\caption{The coding and modulation scheme in the conventional digital system and the designed system.}
\label{F2}
\end{figure*}

\subsection{Non-linear Distortion}
In the pseudo-analog video transmission system, the distribution of the input real-value samples directly determines the spectrum power distribution. However, the hardware can not support unbounded signals in practical due to the limited range of D/A converters. As a consequence, the hardware will truncate signals with high amplitudes, thus introducing the non-linear distortion [32,33]. In order to avoid such non-linear distortion, the transmitted packets are required to be normalized by the application layer, ensuring that the average of the input signals in each packet is equal to 0. Actually, the average does not carry too much information but greatly increases the total transmission power. Besides the normalization, whitening should also be performed on the transmitted signals, which is similar to pseudo-random scrambling and bit stream interleaving in the conventional digital system.

Two whitening methods are adopted in the existing literature of pseudo-analog wireless video transmission system including Hadamard transform [22] and unitary transform [25]. Whitening can bring at least two benefits to the designed  physical layer. Firstly, it ensures that the power spectrum distribution of the transmitted signals is flat so that the transmitted signals can shield the frequency selective fading. Secondly, it reduces the PAPR of AWGN, thus protecting the transmitted signals to suffer from less error caused by the noise.

\section{Transceiver Framework}
In this section, we will introduce the detailed implementations of the designed transceiver framework. Firstly, we will give the overview of the designed transceiver. Then, we will introduce the function of each module contained in the transmitter, channel and receiver, respectively.

\subsection{Overview of the Designed Transceiver}
The designed GNU Radio-based pseudo-analog wireless video transmission transceiver is an open-source real-time signal processing framework, which can support all packet sizes and MCSs (including dense modulation). Fig. 3 shows the detailed implementation block diagram of the transceiver in GNU Radio Companion [34] which is a graphical user interface that can be used to simplify the creation and configuration of flow charts and help researchers understand the data processing process more clearly. Although Fig. 3 does not show the detailed functions of each module, it provides the concept of data flow in GNU Radio. Specifically, the signal processing modules are represented by boxes, and data streams are depicted by arrows connecting signal processing modules. Arrows of different colors represent different data formats, e.g., red for bytes, blue for complexes, and orange for floats. For the ease of understanding, we also provide a simplified transceiver diagram in Fig. 4.
\begin{figure*}[htbp!]
\centering
\includegraphics[width=0.9\textwidth]{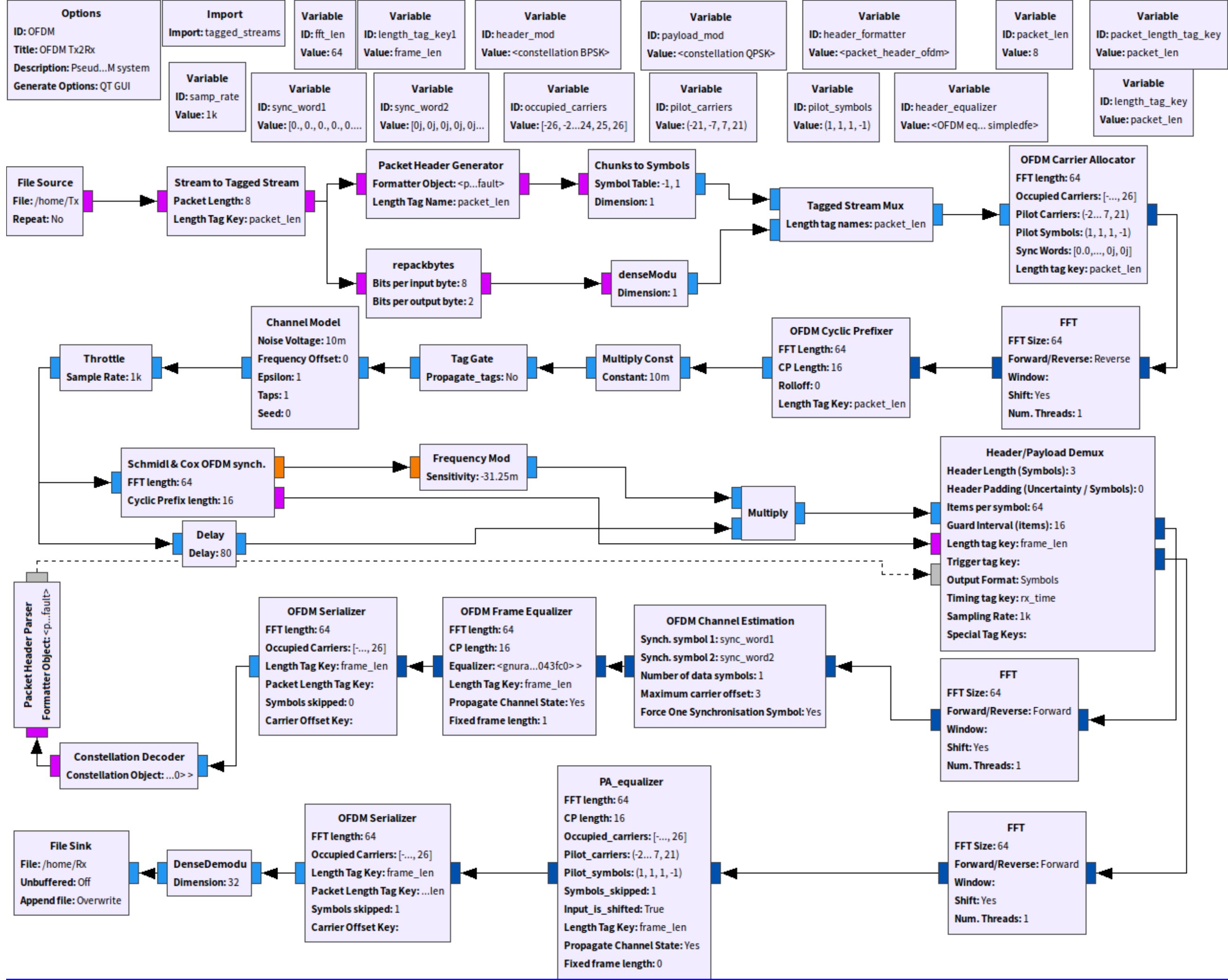}
\caption{The detailed implementation block diagram of the transceiver.}
\label{F3}
\end{figure*}

\subsection{Transmitter}
Compared with the receiver, the transmitter is relatively simpler to implement. As shown in Fig. 4, the input normalized and power-adjusted video data are whitened first. Then, these video data is mapped into complex signals and filled into the frame according to the general frame format as shown in Fig. 5. Specifically, the short training sequence is used for frame detection while the long training sequence is used to determine the start position of the fast Fourier transform (FFT) window. The pilots are used for channel estimation. In the designed system, the payload data represents the modulated complex signals. At the transmitter, we mainly accomplish the following tasks:
\begin{figure*}[htbp!]
\centering
\includegraphics[width=0.9\textwidth]{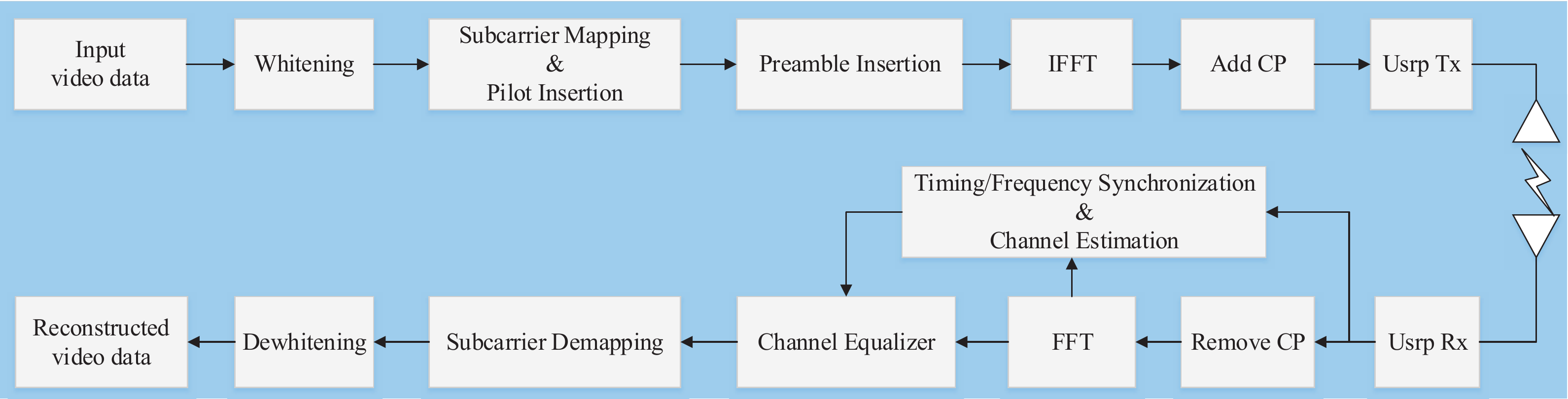}
\caption{The simplified diagram of the transceiver.}
\label{F4}
\end{figure*}
\begin{enumerate}
\item Group DCT coefficients into packets with uniform size and assign a tag containing the length information for each packet.
\item Generate a header (including the short training sequence and the long training sequence) for each packet which can be used for coarse frequency offset tuning at the receiver.
\item Map the packet header to the complex constellation points according to the selected modulation scheme, such as e.g., BPSK, QPSK, 16-QAM, or 64-QAM, and map the payload data to the complex constellation points according to the intensive modulation scheme.
\item Interleave the modulated payload data with pilots.
\item Convert the transmitted signals from frequency domain into time domain using FFT.
\item Add cyclic prefixes to the time-domain signals to deal with inter-symbol interference.
\end{enumerate}
\begin{figure}[htbp!]
\centering
\includegraphics[width=0.5\textwidth]{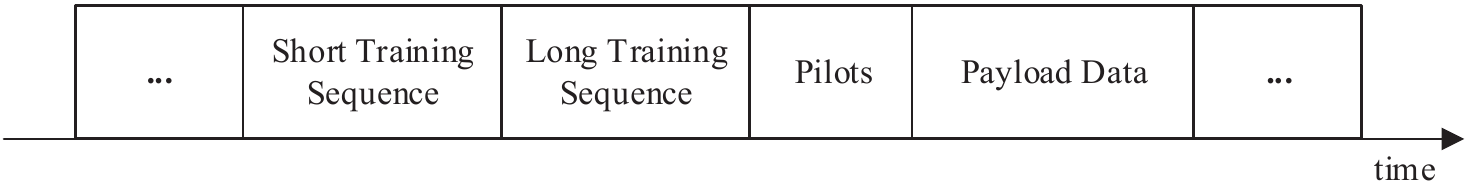}
\caption{The frame format.}
\label{F5}
\end{figure}

The main modules of the transmitter are shown in Fig. 3 and their functions are as follows:
\begin{itemize}
\item Stream to tagged stream: Add tags to the data stream at fixed intervals, which can help the subsequent signal processing modules align data.
\item Packet header generator: Generate a packet header (including the short training sequence and the long training sequence) for the tagged stream which can be used for frame detection and channel estimation at the receiver.
\item Chunks to symbols: Map the bit stream to complex signals according to the selected modulation scheme.
\item Tagged stream mux: Merge multiple tagged streams.
\item OFDM carrier allocator: Generate OFDM symbols according to the frame format.
\item IFFT: Transform OFDM symbols from frequency domain into time domain.
\item OFDM cyclic prefixer: Add cyclic prefixes and perform pulse shaping on OFDM symbols.
\end{itemize}

\subsection{Channel}
One of GNU Radio-based transceiver is that it can provide seamless switch between simulations and experiments. The simulations can be implemented by circulating the transmitted signals back to the receiver without connecting to real radio devices, e.g., USRP. GNU Radio supports researchers to select the simulation environment by providing propagation models such as phase noise, clock drift, AWGN, Rayleigh, and multipath fading. In Fig. 3, we only show the implementation block diagram for simulations. In fact, researchers can switch from simulations to experiments by replacing the channel mode module with the UHD module. The main modules and functions of each module are as follows:
\begin{itemize}
\item Tag Gate: Control whether the tags are transmitted.
\item Channel model: A basic channel model simulator which can be used to help evaluate, design and test various signals, waveforms, and algorithms.
\item Throttle: Ensure that the average rate does not exceed the allowed maximum number of samples per second.
\end{itemize}

\subsection{Receiver}
The design of the receiver is more complicated than that of the transmitter. At the receiver, the decoding of the received signals is mainly divided into two steps: 1) Decode the packet header. The short training sequence and the long training sequence contained in the packet header are used for frame detection and channel estimation, respectively (the channel estimation result will be used for the decoding of payload data); 2)Decode the payload data. After removing the packet headers from the received signals, the remaining payload data is demodulated to the original transmitted signals. The main modules at the receiver and their functions are as follows:
\begin{itemize}
\item Schmidl \& Cox OFDM synch [35]: Use pseudo-random preambles to detect the beginning of a frame and produce fine estimation of carrier frequency offset.
\item Frequency modulator: Convert time-domain signals to frequency-domain signals.
\item Delay: Delay the signals by a certain amount of sampling time.
\item Multiply: Multiply all the input signals.
\item Header payload demux: Demultiplex packet data from burst transmission. When the length of the received packet has not been determined, the packet header is passed to other blocks for demodulation, and then the payload data is demodulated using the information provided by the packet header.
\item OFDM channel estimation: Estimate the channel condition and coarse frequency offset of OFDM with the help of the packet header.
\item OFDM frame equalizer: Equalize the tagged OFDM frames.
\item OFDM serializer: Discard the pilots and extract the payload data.
\end{itemize}

\section{Simulations and Experiments}
In this section, we will analyze the performance of the designed GNU Radio-based pseudo-analog wireless video transmission system. Firstly, we will give the parameter settings of the hardware platform. Then, we will compare the simulation and experiment results from the perspectives of both the objective evaluation metric and the subjective visual quality.

\subsection{Parameter Settings}
The wireless pseudo-analog video transmission system designed in this paper is shown in Fig. 6. The object in the red dashed box represents the Ettus Research X310, which is a universal radio device equipped with two antennas (as shown in the orange dashed box) supporting the full-duplex operation from 120 MHz to 6 MHz in frequency. Ettus Research X310 also has two daughter boards, multiple high-speed interface options (e.g., PCIe, dual 10 GigE, and dual 1 GigE), and a large programmable Kintex-7 FPGA. The open-source software architecture of Ettus Research X310 also supports cross-platform UHD drivers which make it compatible with many development frameworks. In this paper, we run the designed transceiver on a PC (as shown in the green dashed box) which has an Intel i5 CPU with 16 GB RAM. The operating system is Ubuntu 16. 04 with Linux 4. 14. 0. The specific details of the hardware are shown in Table 1.

\begin{figure}[htbp!]
\centering
\includegraphics[width=0.45\textwidth]{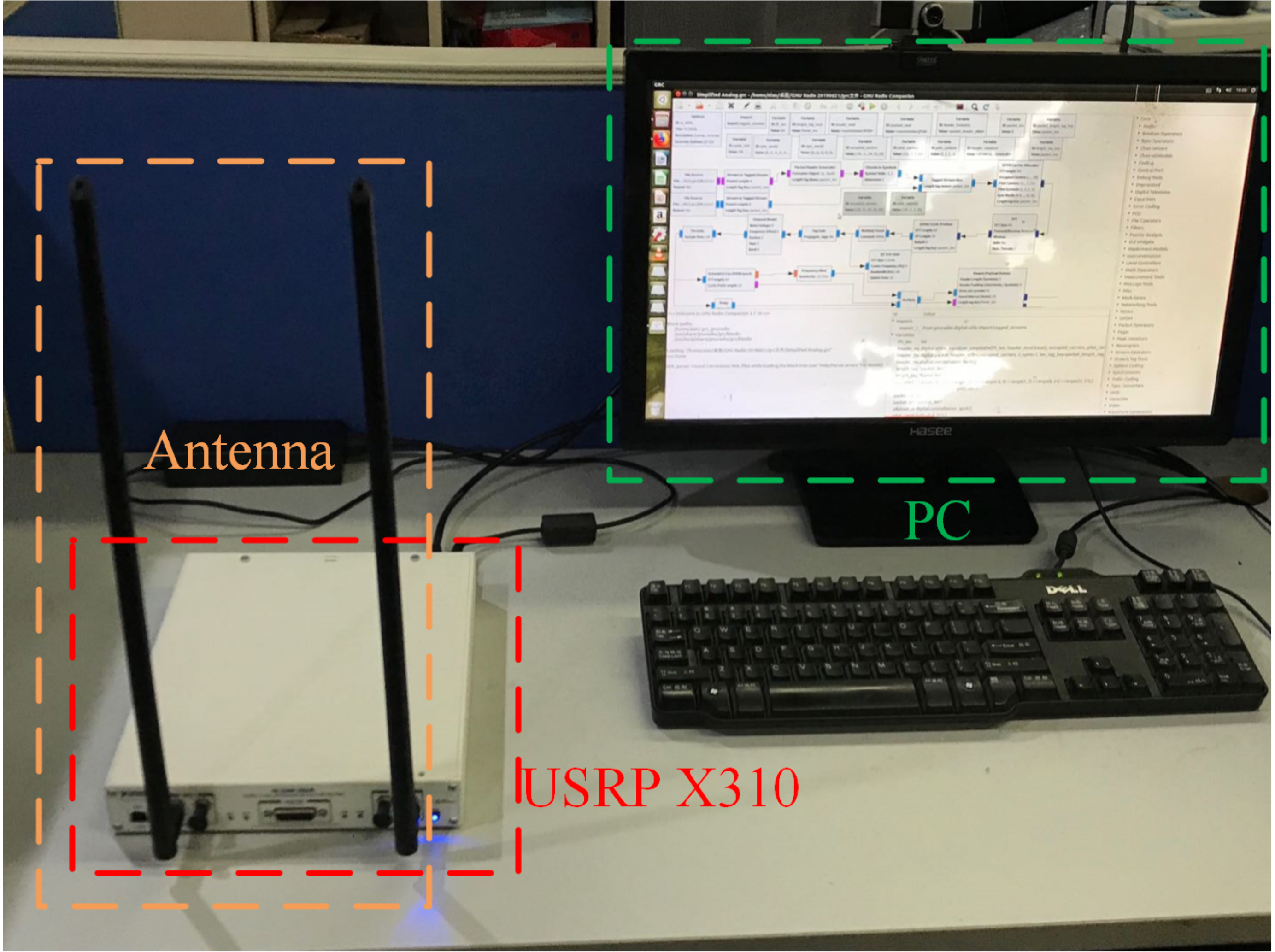}
\caption{The designed GNU Radio-based platform.}
\label{F6}
\end{figure}

\begin{table}[htbp!]
\centering
\caption{The parameter settings.}
\begin{tabular*}{8.3cm}{cc}
\hline
Componets & Types \\
\hline
Operating System & Ubuntu 16.04 with Linux 4. 14. 0  \\
GNU Radio& Version 3.7.10   \\
UHD &Version 3.15.0    \\
SDR & Ettus Research X310   \\
FPGA & Kintex-7  \\
Daughter boards  & UBX-160 \\
Interface  & PCIe/dual 10 GigE/dual 1 GigE  \\
Supported Frequency  & 120 MHz-6 GHz  \\
Maximum Bandwidth  & 20MHz   \\
\hline
\end{tabular*}
\label{T1}
\end{table}

\subsection{Performance Analysis of Simulation and Experiment Results}
In this paper, we use the standard video sequence ``Foreman'' to verify the effectiveness of the designed system [36]. ``Foreman'' has 300 frames with a resolution of $176 \times 144$. Firstly, we test the experiment performance of the designed transceiver at the frequency of 2.4GHz with a channel bandwidth of 20 MHz. Since the USRP radios operate in the same frequency band as 802.11 WLANs, there is unavoidable interference. To reduce the impact of interference, we repeat each experiment ten times. Secondly, we test the simulation performance of designed transceiver over AWGN channel, and set the channel SNR to the same as that of the experiment (please note that the channel SNR can be calculated by the pilots). In addition, we also implement the simulation of video transmission in the conventional digital system under the same channel SNR. Specifically, we generate MPEG4 streams using the H.264/AVC codec provided by the FFmpeg software and the X264 codec library in the conventional digital system. To ensure that they take the same channel bandwidth, the MPEG4 streams are encoded into bit streams at ${1}/{3}$ code rate and mapped into complex signals using 16QAM.

We compare the simulation and experiment results from perspectives of both the objective evaluation metric, e.g., peak signal-to-noise ratio (PSNR) [37] and the subjective visual quality. The PSNR is a standard objective measurement of video/image quality, which can defined as a function of the $M\!S\!E$:
\begin{equation}\label{E1}
P\!S\!N\!R =  20{\log}(\frac{{255}}{{\sqrt {M\!S\!E} }})
\end{equation}
where $M\!S\!E$ represents the mean squared error between all pixels of the reconstructed video and the original version.
\begin{figure*}[htbp!]
\centering
\includegraphics[width=1.0\textwidth]{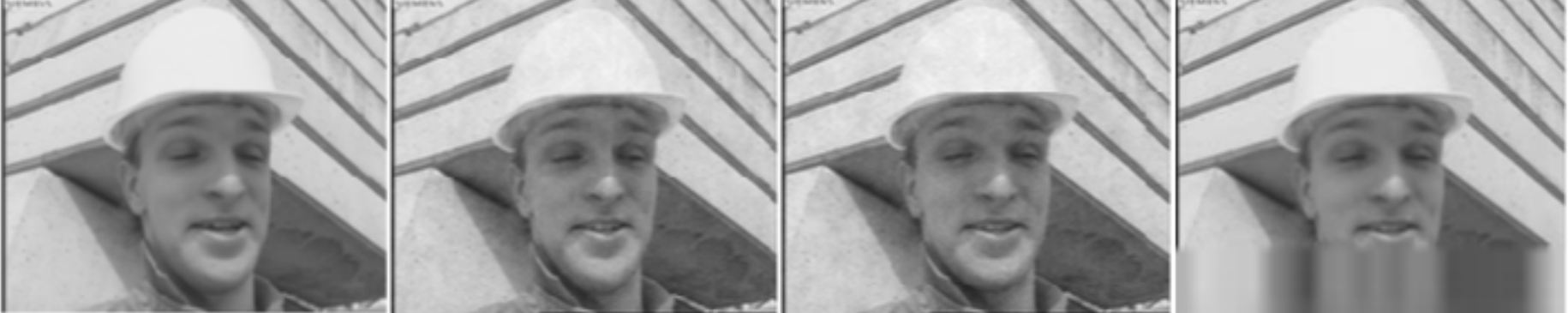}
\caption{Simulation and experiment results (Foreman \#2). From left to right: (a) the original frame; (b) the reconstructed frame using the designed transceiver by experiment, PSNR = 28.71dB; (c) the reconstructed frame using the designed transceiver by simulation, PSNR = 28.95dB; (d) the reconstructed frame using digital transmission system by simulation, PSNR = 23.52dB.}
\label{F7}
\end{figure*}

\begin{figure*}[htbp!]
\centering
\includegraphics[width=1.0\textwidth]{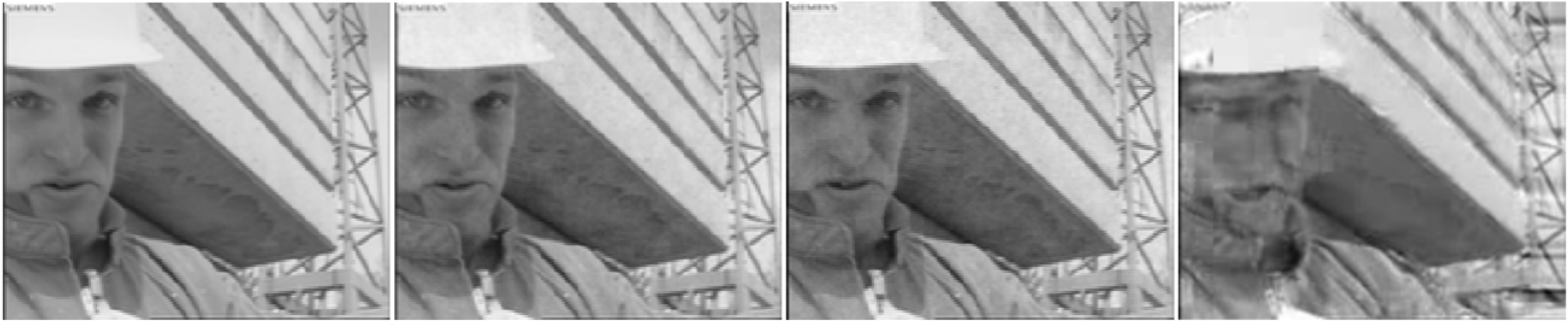}
\caption{Simulation and experiment results (Foreman \#180). From left to right: (a) the original frame; (b) the reconstructed frame using the designed transceiver by experiment, PSNR = 27.14dB; (c) the reconstructed frame using the designed transceiver by simulation, PSNR = 27.36dB; (d) the reconstructed frame using digital transmission system by simulation, PSNR = 21.11dB.}
\label{F8}
\end{figure*}
The simulation and experiment results are shown in Figs. 7-8. For simplicity, we select the $2^{nd}$ frame  (see Fig. 7) and the $180^{th}$ frame (see Fig. 8) of ``Foreman'' to analyse the performance of the designed transceiver. In Figs. 7-8, we present four images from left to right including 1) the original frame, 2) the reconstructed frame using the designed transceiver by experiment, 3) the reconstructed frame using the designed transceiver by simulation, 4) the reconstructed frame using the digital system by simulation.

From Figs. 7-8, one can see that the designed system can achieve the similar performance in the simulation and experiment in terms of subjective visual quality. Since the thermal noise caused by the actual radio device may affect the quality of the reconstructed frame, the PSNR performance of the simulation results are slightly higher than that of the experiment results using the designed transceiver. One can also conclude from Figs. 7-8 that pseudo-analog video transmission system can obtain better performance compared with the digital video transmission system. Although the frames reconstructed by the pseudo-analog system are a little vague compared with the original frames, one can also receive the most information contained in the frames. However, in the same channel condition, the reconstructed frames by the digital system are prone to distortion which are not suitable for watching. This is because the digital system adopts the inter-frame compression and motion compensation technologies, resulting in a high correlation between frames. The demodulation error of a single frame may result in the loss of several-second video clip.

\section{Conclusions}
In this paper, a GNU Radio-based pseudo-analog video transmission transceiver has been proposed which can
provide seamless switch between simulations and experiments. Firstly, we have described the fact that pseudo-analog transmission system can also achieve the same optimal performance as digital system from the perspective of rate-distortion theory. Then, we have analyzed the two difficulties in the practical implementation process including data format modification and non-linear distortion. Next, we have provided  the detailed flow chart of the proposed transceiver and introduced the detailed function of each module. Finally, we compare the simulation and experiment results in terms of PSNR and the subjective visual quality. The testing results have shown that the designed transceiver can achieve the similar performance in the simulation and experiment which shows the effectiveness of the designed transceiver. In addition, we have also compared the simulation and experiment results obtained by the designed transceiver with the simulation results obtained by the digital system. The simulation results have indicated that the pseudo-analog system can achieve over 6.25dB PSNR gains than the digital system under the same channel condition.

\section{Acknowledgement}
This work was supported in part by the National Natural Science Foundation of China under Grants No. U1733114 and No. 61631017, in part by the Fundamental Research Funds for the Central Universities, in part by Shanghai Rising-Star Program under Grant No. 19QA1409100, and in part by Shanghai Science and Technology Innovation Action Plan under Grant No. 19DZ1201100.

\begin{IEEEbiography}[{\includegraphics[width=1in,height=1.25in,clip]{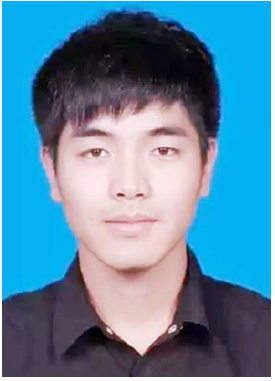}}]{Xiao-Wei Tang} (\emph{S'16, IEEE}) received the B.E. degree in Communication Engineering from Tongji University in 2016, where he is currently pursuing the Ph.D. degree. He has published several research papers on IEEE Transactions on Multimedia, IEEE Access, IEEE Globecom, and Mobile Networks \& Applications. He was a recipient of the Excellent Bachelor Thesis of Tongji University in 2016, the National Scholarship for Graduate Students by Ministry of Education of China in 2017, the Outstanding Students Award of Tongji University in 2017, the Outstanding Freshman Scholarship of Tongji University in 2018, the Chinese Government Scholarship by China Scholarship Council in 2019, the Outstanding Students Award of Tongji University in 2019, and the National Scholarship for Graduate Students by Ministry of Education of China in 2019. From Aug. 2019, he is doing research on UAV-enabled wireless video transmission in the Department of Electrical and Computer Engineering, the National University of Singapore, as a visiting scholar. His research interests include pseudo-analog video transmission, UAV communication, convex optimization, and deep learning.
\end{IEEEbiography}
\begin{IEEEbiography}[{\includegraphics[width=1in,height=1.25in,clip]{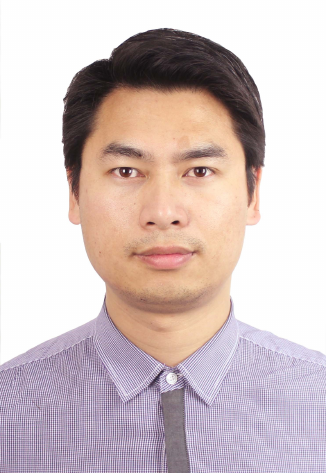}}]{Xin-Lin Huang}
(\emph{S'09-M'12-SM'16, IEEE}) is currently a professor and vice-head of the Department of Information and Communication Engineering, Tongji University, Shanghai, China. He received the M.E. and Ph.D. degrees in information and communication engineering from Harbin Institute of Technology (HIT) in 2008 and 2011, respectively. His research focuses on Cognitive Radio Networks, Multimedia Transmission, and Machine Learning. He published over 70 research papers and 8 patents in these fields. Dr. Huang was a recipient of Scholarship Award for Excellent Doctoral Student granted by Ministry of Education of China in 2010, Best PhD Dissertation Award from HIT in 2013, Shanghai High-level Overseas Talent Program in 2013, and Shanghai Rising-Star Program for Distinguished Young Scientists in 2019. From Aug. 2010 to Sept. 2011, he was supported by China Scholarship Council to do research in the Department of Electrical and Computer Engineering, University of Alabama (USA), as a visiting scholar. He was invited to serve as Session Chair for the IEEE ICC2014. He served as a Guest Editor for IEEE Wireless Communications and Chief Guest Editor for International Journal of MONET and WCMC. He serves as IG cochair for IEEE ComSoc MMTC, and Associate Editor for IEEE Access. He is a Fellow of the EAI.
\end{IEEEbiography}

\vfill

\end{document}